# Confusion in Electromagnetism and Implications of CPT Symmetry
## - System of Units Associated with Symmetry -

Tsuneo Ichiguchi
*Information and Communications Research Unit*
National Institute of Science & Technology Policy,
Ministry of Education, Culture, Sports, Science & Technology

## 1 │ Introduction

The year 2008 was a year that Japan impressed the world with its high level of science, with four Nobel Prize winners who were either Japanese or of Japanese origin. This Nobel Prizes also taught us the importance and potential of basic science. Basic science is not just important as the common knowledge of humankind. Persons with the scientific mind can play an active role in technological innovation beyond a mere extension of the present. History shows that physics, one of basic science, experienced three revolutions in the late 19th and early 20th century, namely, Maxwell and Hertz's electromagnetism, Einstein's theory of relativity, and quantum mechanics developed by Planck, Bohr, de Broglie, Shrödinger, Heisenberg, and Dirac. It goes without saying that many engineering fields, as well as physics, are still based on electromagnetism, the theory of relativity, and quantum mechanics. They are not simply academic disciplines; they significantly benefit our lives as well. For example, there would be no radio, TV, or cellular phone if we had not come to know of the existence of electromagnetic waves. Without the knowledge of quantum mechanics, the operating mechanism of semiconductor devices would be incomprehensible; there would be no computers. The theory of relativity has a key role in the definition of units of time and length.

The CPT symmetry, which was applied to elementary particles by Kobayashi and Masukawa, is an important concept. C-symmetry (charge symmetry) refers to the symmetry between particles and antiparticles. P-symmetry is the symmetry under space reversal. P-symmetry is also referred to as parity or chirality, which is the symmetry of the left-handed and right-handed coordinates. T-symmetry is the symmetry under time reversal. Associating the universal constants, such as the speed of light $c$, permittivity $\varepsilon_0$ or permeability $\mu_0$ of free space, the charge $e$ of electrons, and the Planck constant $h$, with reversal symmetries, we can reveal the profound meaning of the constants. It would be quite important to the coming revolution. In the International System of Units (or MKSA units), meter, kilogram, second, and ampere are the fundamental units, which are defined for human convenience without any universal meaning. The questions of why the light of speed has a constant value of approximately 300,000 km/s and why each individual electron has exactly the same charge of -1.602773 × 10$^{-19}$ Coulombs are thus unexplainable. However, by adopting a system of units where $c^2 = \varepsilon_0{}^2 = e^2 = h^2 = 1$, and by associating the constants with CPT symmetry, we can answer the questions.

The electromagnetism, established by Maxwell, is essential in a variety of fields, such as physics, electronics, electrical engineering, and material science. There is confusion, however, in standard textbooks of electromagnetism as to whether to support *E-B* formulation or *E-H* formulation. There are two types of magnetic fields, *H* (magnetic field) and *B* (magnetic induction or magnetic flux density), and the difference standpoints exist; which field is more fundamental. One of the reasons for this confusion is an insufficient study of the space reversal symmetry in modern textbooks, although Maxwell pointed out the significance of P-symmetry in his textbook in the 1870s. We can reveal the meaning of $\varepsilon_0$ and $\mu_0$ through a discussion of P-symmetry.

While introducing the "confusion" in electromagnetism, we also would like to emphasize here that we need scientists who persevere in pursuing such fundamental problems.

## 2 | Time, Space, and Relativity - In the beginning there was light-

### 2-1 On time reversal symmetry

In CPT symmetry, T-symmetry is the symmetry under time reversal. Equations of motion, such as Newton's equations, Maxwell's equations, and Shrödinger's equations, always include time $t$ as a parameter. The reversing of time refers to the transformation from a system where the parameter $t$ flows forward to a system where $t$ is reversed ($t' = -t$). If the equation remains exactly the same regardless of whether time is reversed or not, the equation will be referred to as conserving (or satisfying) the time reversal symmetry. Newton's equations and Maxwell's equations conserve the time reversal symmetry. Assuming that the sign of the imaginary unit $i$ changes under time reversal, Shrödinger's equation also conserves the time reversal symmetry. Newton's, Maxwell's, and Shrödinger's equations are all differential equations, and if the equations are integrated (for example, the parabolic curve of a free fall) they also satisfy the time reversal symmetry. In discussions of time reversal symmetry, simple conversion of an increase in time into a decrease in time (i.e., playing a film in reverse) is insufficient; we must always consider the situations where the forward flow of time is switched to the reverse flow of time, and where the future is switched to the past.

When the direction of time is reversed, some quantities do not change the sign (or direction) like position vector $x(t) \rightarrow x(t') = x(-t)$, while some quantities change the sign (or direction) like velocity $v(t) \rightarrow -v(t') = -v(-t)$. This will be referred to as parity under time reversal. Other quantities that do not change the sign or direction are area, volume, acceleration, force, electric field, voltage, electric resistance, and energy. Quantities that change the sign or direction under time reversal are time itself, time derivatives, momentum, angular momentum, electric current, and magnetic field. We can easily determine the time reversal symmetry of an equation, if the symmetry that changes sign (or direction) under time reversal is denoted by

T(-) and the symmetry that does not change sign by T(+). The values of T(-) and T(+) are -1 and +1, respectively. By definition, T(+) = T(+)・T(+), T(-) = T(+)・T(-), and T(+) = T(-)・T(-). The symbol T is added for the sake of distinction from C-symmetry and P-symmetry. The dot mark (・) represents an ordinary multiplication, and is used to make a distinction from a cross product to be described later. For example, symmetry T(+) = T(-)・T(-) holds for the equation $x$ (distance) = $v$ (velocity)・$t$ (time), showing that time reversal symmetry is conserved. Thus the symmetry of an equation is conserved when the symmetries of the left side and the right side are the same.

An equation always conserves time reversal symmetry, with one significant exception. Time reversal symmetry breaks when entropy (or randomness) increases or when energy is dissipated in the form of heat: the examples are Ohmic law and air friction acting in proportion to velocity. In Ohmic law $V$ (voltage) = $I$ (current)・$R$ (resistance), current is reversed under time reversal, while voltage and resistance are not. Thus, equation of symmetry is T(+)≠T(-)・T(+). If time reversal symmetry of resistance is T(-), the resistance that had been consuming energy through heat generation would in turn generate electricity, which violates the laws of thermodynamics. Consequently, the time reversal symmetry of resistance should be T(+), and symmetry of the left side of Ohmic law differs from the right side. Such a situation is referred to as time reversal symmetry being "broken" or "not conserved." The breaking of time reversal symmetry always follows an entropy increase.

### 2-2 History of One Second

One second is the fundamental unit of time in physics and engineering. The unit of time used to be determined from a single day on the Earth or from the rotation period. Historically (from the 1930s to 1956), the second was defined in terms of the rotation of the Earth as 1/86,400 of a mean solar day. The period of the Earth's rotation was then found to slightly fluctuate because of tides and other factors and to become longer as well. The definition of the second was thus changed to an ephemeris time (ET) based on the period of revolution of the Earth around the Sun, i.e., one second is 1/31,556,925.9747 of the solar year, which was used from 1956 to 1967.

The reason for this is that the period of revolution is more stable than that of rotation. Incidentally, the length of a single day 500 million years ago is estimated to be approximately 21 hours.[1]

Subsequently, with the development of atomic clocks, the definition of the second was changed to atomic time no longer based on the Earth's rotation or revolution; i.e., one second is the duration of 9,192,631,770 periods of radiation corresponding to the transition between two hyperfine levels of the ground state of the cesium-133 atom. This definition in atomic time has hitherto been used. Using the periods of electromagnetic waves (microwaves) generated by the cesium-133 atom as the reference does not necessarily imply that a single day of the Earth no longer serves as the standard. If we are indeed going to adopt atomic time based on the periods of electromagnetic waves, one second might be defined not as 9,192,631,770 periods but as a round number of 10,000,000,000 periods. However, such a definition results in a discrepancy from the Earth's rotation time and is inconvenient for daily life, which is why the odd figure of 9,192,631,770 periods has been used.

Despite the adoption of atomic time, time standards are still based on the rotation (to be precise, one day) of the Earth. Since the Earth's rotation is gradually slowing down and one second is gradually getting longer, it is necessary to "adjust" atomic time. A "leap second" is thus inserted for the sake of adjustment. On January 1, 2009, a leap second adjustment was carried out by inserting "8:59:60" between 8:59:59 a.m. and 9:00:00 a.m.[2] Twenty-four leap seconds have already been inserted since 1972. The reason for the adjustment is that we essentially use the Earth's rotation time. With the Earth's rotation slowing down, it will be necessary to insert leap seconds more frequently in the future.

As has been described above, even though the second is one of the fundamental units in physics and engineering, it is an artificial (and arbitrary) unit based on the Earth's rotation for the sake of human convenience. The second is therefore not a unit of time suited to describing universal truths or phenomena.

## 2-3 History of One Meter

Space is measured with unit of length, and we use a meter as the base unit of length. The meter was originally defined as one ten-millionth of the Earth's meridian from the North Pole to the Equator through Paris, from which value several platinum-iridium standard bars were made. Today, it is known that the exact distance from the North Pole to the Equator is 10,002.288 kilometers and the Earth is slightly oblate. It is not accidental that the Earth has a circumference of almost exactly 40,000 kilometers.

The standard meter bar may vary in length depending on temperature (thermal expansion), corrosion, or other reasons. In 1960, at the 11th General Conference of Weights and Measures, the meter was thus redefined as 1,650,763.73 wavelengths of light corresponding to the transition between the 2p10 and 5d5 levels of the krypton-86 atom. The idea of using the wavelength of light as the unit of length had already appeared in Maxwell's writings in the 1870s, but took almost 90 years to realize.

It was at the 17th General Conference of Weights and Measures in 1983 that the definition of length changed in essence; one meter is the distance traveled by light in vacuum during a time interval of 1/299,792,458 of a second. To be precise, however, length was not defined directly, but rather the speed of light was defined as $c = 299,792,458$ m/s. Length is not defined until time, i.e., 1/299,792,458 of a second, is determined. In other words, length is defined by time. It follows that the speed of light is no longer a quantity to be measured by experiment but to be assigned by definition. We can see this as a step toward "the principle of constancy of the speed of light" which is one of the fundamentals of Einstein's theory of special relativity. The constancy of the speed of light has been proven by experiment, whereas the physical reason, why the speed of light should be constant, remains unexplained. Einstein called it "principle" because it was a "correct but theoretically unprovable hypothesis."

The current definition of length, even using the universal quantity of the speed of light, is based on the size of the Earth, which is not universal from a cosmic point of view. In fact, length is defined using the odd figure of 1/299,792,458 of a second, but not using one second. The original idea of the metric system that depends on the Earth's scale still survives, and we can never understand the meaning of defining length in terms of the speed of light nor enjoy the essential merits. The metric system of units currently in use is only significant as a world common language. In terms of physics, the base units of seconds, meters, kilograms, and amperes thus have no universal meaning.

## 2-4 Why the Speed of Light is Constant

Length is defined by time using the speed of light because time and length can no longer be defined independently. Using the speed of light $c$, the distance $x$ for light to travel is expressed as:

$$x \text{ (distance)} = c \text{ (speed of light)} \cdot t \text{ (time)}. \qquad (1)$$

This relation shows that time and length (or distance) are not independently determinable. Length is determined once time is set, and time is determined once length is set. That is, we can only set either time or length freely, but not both of them at once.

The unit of the "light-year" used in astronomy represents the distance that light travels in one year. The light-year expresses length in terms of time. Using this terminology, we can say that the current definition of one meter is 1/299,792,458 light-seconds. A light-second is a more natural unit of length to take unless we do focus on the size of the Earth. One light-second is equal to approximately 300,000 kilometers. If the figures are too high, we can use nano-light-second, which is 0.299792458 meters or approximately 30 centimeters. The scale may be suitable for our daily life because it close to one shaku (= 0.303 meters) or one foot (= 0.3048 meters). Conversely, time can be defined in terms of length. For example, "a light-meter" may be the time necessary for light to travel one meter. This, however, complicates correspondence with the Earth's rotation time. Such a unit might become necessary in the distant future, but for now it would be more convenient to adopt the light-second as the base unit of length, leaving time intact.

The use of the light-second as the base unit of length means that the speed of light is defined as one. In physics, we sometimes use natural units where $c = 1$, the essential meaning of which is that length and time are the same dimensions. Accordingly, velocity is a dimensionless value and has no unit. Velocity can be expressed as 0.1 times or 0.00001 times the speed of light, for instance, so that velocity can be described without definitions of time and length. The International System of Units (or MKSA system of units) currently in use is a system with four base units; meters, kilograms, seconds, and amperes. When time and space are unified by $c = 1$, the unit of length (or unit of time) disappears, resulting in a system with three base units.

When an entropy does not increase, time reversal symmetry must be conserved in any and all equations, including equation (1). The time reversal symmetry of distance $x$ is T(+) and that of time $t$ itself is T(-). In order for equation (1) to conserve time reversal symmetry, the speed of light $c$ should therefore have symmetry T(-). That is, $c = -1$ in the time-reversed world. It is natural for the sign to change since the speed of light represents a velocity. Since $c^2 = 1$ holds even in such a case, equation (1) can be rewritten as:

$$t \text{ (time)} = c \text{ (speed of light)} \cdot x \text{ (distance)}. \qquad (2)$$

This equation holds because the speed of light has no unit, and because time and distance are expressed in the same unit. The speed of light $c$ is either +1 or -1, which is yet to be determined and thus cannot be written numerically.

Equation (1) suggests that time multiplied (or operated) by the speed of light becomes distance. Equation (2) suggests that distance multiplied (or operated) by the speed of light becomes time. This reveals that the speed of light $c$ is a quantity (or operator) that transforms time into space and space into time. Equation $c^2 = 1$ is the condition under which time and space maintain their original scale without expansion or contraction after "time"→"space" →"time" transformations or "space"→"time"→"space" transformations. Such a condition can be said to provide the exact explanation for the principle of the constancy of the speed of light. It also implies that space is curved under $c^2 \neq 1$ (the general theory of relativity).

Multiplying (or operating) a certain equation by the speed of light $c$ (= ±1) changes the time reversal symmetry of the equation. The speed of light $c$ can thus be interpreted as an operator for changing time reversal symmetry or as an operator for time reversal. Here, $c^2 = 1$ is the condition under which time maintains it's original scale. As described so far, the universal system of units where $c^2 = 1$ is important in discussing the symmetry of space and time. Such a system of units even includes the essence of the theory of relativity.

## 3 | Application of Space Reversal Symmetry to Electromagnetism

### 3-1 On Right-handed and Left-handed Relations in Space

Spatial translational symmetry and rotational symmetry have significance for crystal engineering, semiconductor physics, quantum mechanics, etc. This report will not deal with translational symmetry and rotational symmetry, however, but only with space reversal symmetry. Space reversal symmetry refers to P-symmetry within CPT symmetry, and is also called parity. It signifies the symmetry of transformation between right-handed coordinate system and left-handed coordinate system.

We usually use a "rectangular coordinate system" (Cartesian coordinates) which is defined by three mutually orthogonal reference lines, $x$, $y$, and $z$ axes in a three-dimensional space. The $x$, $y$, and $z$ axes are collectively referred to as coordinate axes, which we can chose in two ways. One is to associate the coordinate axes with the left thumb and fingers as shown in Figure 1(a). The other is to associate the coordinate axes with the right thumb and fingers as shown in Figure 1(b). The resultant systems are called the "left-handed coordinate system" and

"right-handed coordinate system", respectively. The left-handed coordinates can be strictly distinguished from the right-handed coordinates since a rotation of the coordinates will not transform one into the other.

Although the right-handed coordinate system is used traditionally as a rule, laws of nature never choose one out of the right-handed and left-handed coordinates. The symmetry of the right and left hands is also referred to as chiral symmetry. Some biogenic substances, such as amino-acids produced by living matter, are known to choose the left-handed symmetry. Physical laws, however, have the same expressions in both systems of the right-handed and left-handed coordinates, with one exception of beta decay (or weak interactions).

The two coordinate systems shown in Figures 1(a) and 1(b) have the $z$ axis in common, with the $x$ and $y$ axes exchanged, so that the axes are not in equivalent positions. To flip from the right-handed to the left-handed coordinates with the equivalent position of three axes, all three axes are reversed (Figure 2). Suppose now that vectors, or quantities that have both length and direction, are transformed from the right-handed to the left-handed coordinates; see the thick arrows in Figure 2. There are two types of vectors, ones that are transformed in the same direction and others that are transformed in the reverse direction. The ones transformed in the same direction are called axial vectors (or pseudovectors). The ones transformed in the reverse direction are called polar vectors (or true vectors). Examples of polar vectors are a position vector $\boldsymbol{r}$, velocity $\boldsymbol{v}$, acceleration *a*, momentum *p*, force *F*, and electric field *E*. Examples of axial vectors are angular momentum *l*, torque *N*, and magnetic induction *B*. Many of the axial vectors are relevant to rotation in some way.

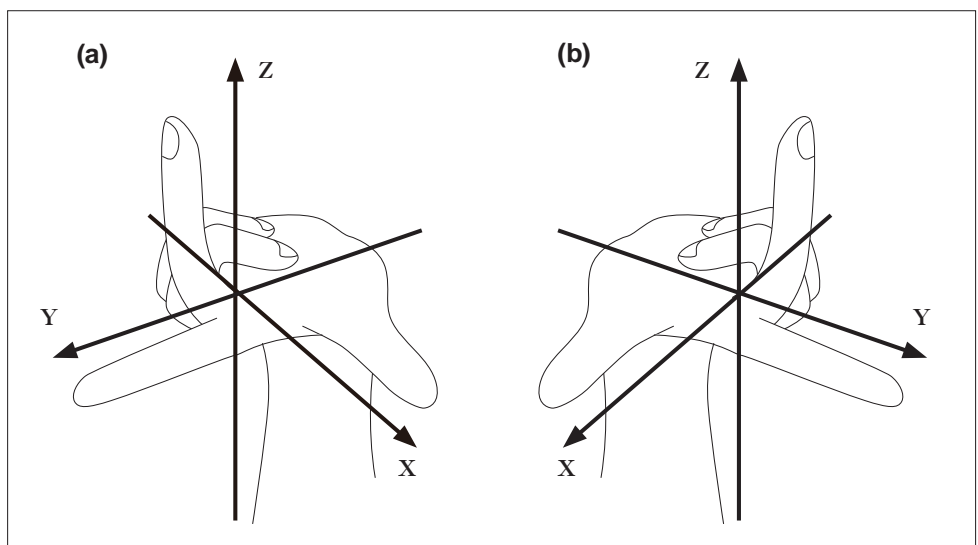

**Figure 1 :** (a)the left-handed and (b) right-handed coordinate systems

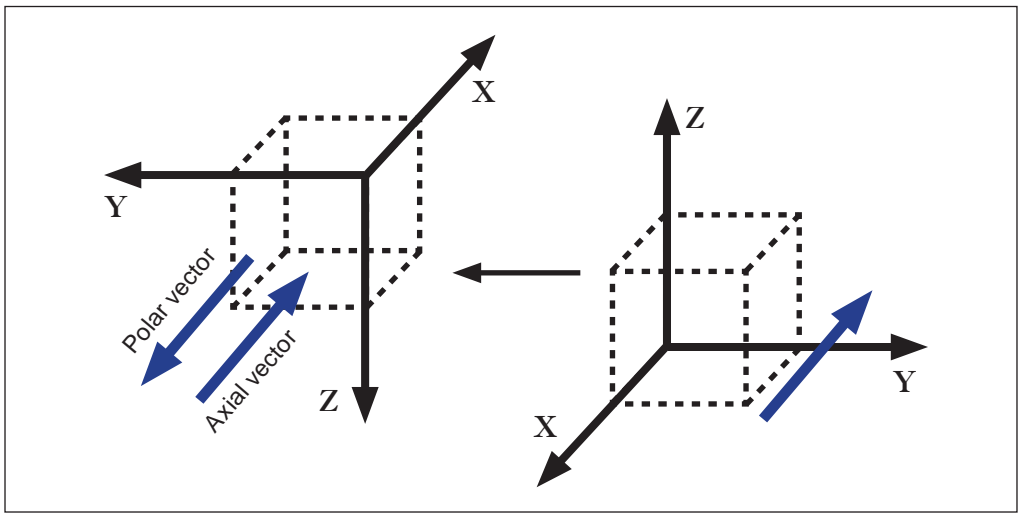

**Figure 2 :** How vectors are transfered from the right-handed to left-handed coordinates

The distinction between polar vectors and axial vectors appears even in ordinary textbooks on vector algebra. One of the examples reads: "If a polar vector such as velocity is reflected in a plane mirror perpendicular to its direction, the direction of the velocity appears to reverse. On the other hand, if an axial vector such as angular velocity is reflected in a mirror perpendicular to the vector, the angular velocity in the image remains unchanged in the direction of rotation."[3] In fact, the direction of velocity appears to reverse when reflected in a plane mirror perpendicular to its direction. It should be noted, however, that the coordinate axis of that direction also reverses. The description that an axial vector remains unchanged while a polar vector changes it's direction under the transformation is nothing but a case of the left-handed coordinates being seen from the viewpoint of the right-handed coordinates. In other words, the image in the mirror is viewed and described from outside the mirror, which is not correct reversal of space. As seen in Figure 2, the fact is that space itself is inverted, or equivalently, the coordinate axes are reversed. With this in mind, we can conclude that the direction of polar vectors remains unchanged whereas the direction of axial vectors reverses. A clearer explanation could be given by using mathematical expressions with unit vectors along the coordinate axes and vector components, but this is omitted here to avoid complexity.

The space reversal does not change the direction of polar vectors but it reverses the direction of axial vectors. When the sign or direction does not change under space reversal, we will denote the symmetry by P(+). When the sign or direction changes under space reversal, we will denote the symmetry by P(-). We assign +1 and -1 to P(+) and P(-), respectively. P(+) and P(-) can be used in the same way as the foregoing T(+) and T(-). An equation where the left-hand side and right-hand side coincide on the symmetry is referred to the symmetry being satisfied or conserved. The symbols P(+) and P(-) are useful in clarifying the space reversal symmetry and in checking the conservation of the symmetry. All equations, including Newton equation, Maxwell equations, and Shrödinger equation, are known to conserve the space reversal symmetry. One and only exception is the equation that include weak interactions. However, there is confusion in Maxwell equations as will be described later. That is, a confusion about space reversal symmetries of magnetic field *H* and electric induction field *D*.

It is well known that the cross product of two polar vectors is an axial vector, like angular momentum *l* = *r*(position vector) × *p*(momentum) and torque *N* = *r*(position vector) × *F*(force), where *r*, *p*, and *F* show the symmetry P(+). Given that the cross product "×" shows symmetry P(-), the foregoing two equations have symmetry P(-) = P(+) · P(-) · P(+), thus conserving symmetry. We give the symmetric symbol P(-) to the cross product × itself, because it differs from dot product (or scalar product) of two vectors and from ordinary multiplications in terms of space reversal symmetry. This is related to the fact that cross product changes direction if the order of multiplication is reversed, while dot product and ordinary multiplication maintain the same sign even if the order is reversed. In this report, we will use the symbol "×" for cross product and the symbol "・" for both dot product and ordinary multiplication. Using P(+) and P(-), we can easily deduce that a cross product of polar vector and axial vector is a polar vector. An example of this is the Lorentz force, which is given by a cross product of velocity $v$ and magnetic induction $B$. When we consider rot $A$ (or $\nabla \times A$ ) which is the rotation of vector $A$, the derivative operator "rot" or " $\nabla \times$ " has the symmetry P(-). An example of this is the relation between vector potential $A$ and magnetic induction $B$ (i.e., $B = \text{rot} A$).

Under space reversal, some quantities change sign even though they have no spatial direction. An example of this is the triple scalar product $V = (a \times b) \cdot c$. As shown in Figure 3, this scalar product represents the volume of the parallelepiped (or rectangular parallelepiped) formed by the coterminous sides $a$, $b$, and $c$. The triple scalar product is sometimes written simply as [$abc$] since it remains the same even if the order of the three vectors is cyclically shifted. Replacing $a$ with $b$ to give [$bac$], however, changes the sign. This means that triple scalar products in the left-handed and right-handed coordinates differ in the sign, and they show the symmetry P(-). This is because the triple scalar product contains a cross product. "Such a scalar that changes sign depending on the configuration of the coordinate axes is named pseudoscalar."[3] In contrast, a scalar that does not change sign irrespective of the configuration of the coordinate axes is referred to as true scalar or simply as scalar. Pseudoscalars and true scalars show the symmetries P(-) and P(+), respectively.

The foregoing discussion implies that volume is defined as a negative quantity in the left-handed coordinates although it is a positive quantity in the right-handed coordinates, which was pointed out by Maxwell in his famous textbook on electromagnetism, *A Treatise on Electricity and Magnetism*.[4] The textbook contains a section titled "On Right-handed and Left-handed Relations in Space", and it reads: "*This relation between the two (cross) products dx×dy and dy×dx may be compared with the rule for the product of two perpendicular vectors in the method of Quaternions, the sign of which depends on the order of multiplication; and*

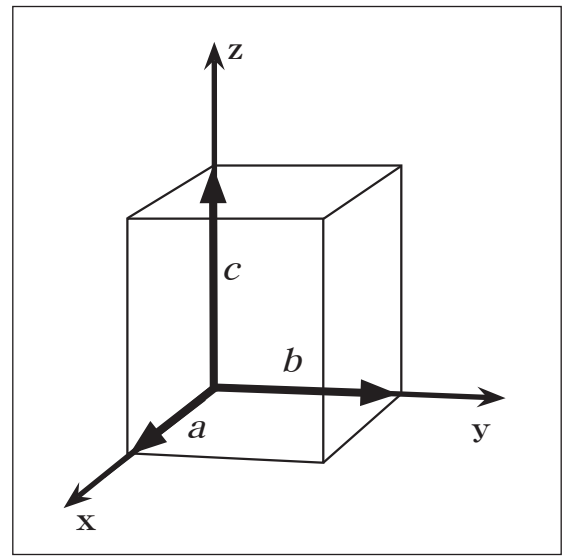

**Figure 3 :** A triple product of three vectors $a$, $b$, and $c$ represents the volume of a parallelepiped (or rectangular solid)

*with the reversal of the sign of a determinant when the adjoining rows or columns are exchanged. For similar reasons a volume integral is to be taken positive when the order of integration is in the cyclic order of the variables x, y, z, and negative when the cyclic order is reversed.*" Here, the term determinant has the same meaning as triple scalar product because the product is equivalent to the determinant when components of vectors are arranged properly. As Maxwell pointed out, volume is negative in the left-handed coordinates. In the following section, the implication of negative volume will be discussed in more detail.

## 3-2 Negative Volume in Left-handed Coordinates

There are only two textbooks that clearly state that volume is negative in the left-handed coordinates: the one written by Maxwell himself and *Mathematics in Classical Physics*[5] by Isao Imai. The latter is based on Maxwell's textbook and it is not surprising that it states that volume is negative in the left-handed coordinates. Imai's textbook also gives us clues about what Maxwell thought of electric and magnetic fields. Shigeo Kobata's *Thus were Created the Electromagnetic Units*[6] points out that the confusion in post-Maxwell electromagnetism started with Sommerfeld's textbook.

It is commonly understood that the triple scalar product of three vectors is a pseudoscalar and thus changes sign depending on the left-handed and right-handed coordinates, and none would argue against it. The question is whether to take an absolute value of the product and to define volume as positive all the time. If we were to take only the right-handed coordinates and not the left-handed coordinates, volume is always positive and this problem would not occur. The problem is unavoidable, however, since space reversal symmetry refers to a transformation from the right-handed to the left-handed coordinates or from the left-handed to the right-handed coordinates. The artificial operation of taking the absolute value to keep volume positive all the time would bring discontinuity and inconsistency in the description of space. In the following discussion, volume in the left-handed coordinates will thus be considered negative as Maxwell intended.

What is the implication of negative volume in the left-handed coordinates? Density, defined by dividing the mass of a body by the volume it occupies, becomes negative in the left-handed coordinates. Negative mass is problematic, but negative density is not if it is defined so. The symmetry of the equation $M(\text{mass}) = \rho(\text{density}) \cdot V(\text{volume})$ is P(+) = P(-)·P(-), where the space reversal symmetry is conserved. Similarly, energy is positive and energy density is negative in the left-handed coordinates. Charge and charge density also have opposite signs. Mass, energy, and charge have the space reversal symmetry P(+) while density, energy density, and charge density P(-).

Since the symmetry of charge density is P(-), that of current density is also P(-). Maxwell equations include both charge density and current density that may affect the space reversal symmetry of the equations. When volume is negative in the left-handed coordinates, resultant symmetries of electromagnetic fields are as follows: electric field $E$ and magnetic field $H$ are polar vectors with symmetry P(+), and electric induction $D$ (or electric flux density) and magnetic induction $B$ (or magnetic flux density) are axial vectors with symmetry P(-). The conclusion (see Table 1) was previously stated in the textbook of *Mathematics in Classical Physics* mentioned above. Maxwell himself made a clear distinction between them. He considered $E$ and $H$ as quantities defined with respect to a line, and $D$ and $B$ as quantities defined with respect to a plane.[4,6] This suggests that Maxwell himself thought of $E$ and $H$ as polar vectors and $D$ and $B$ as axial vectors.

On the other hand, starting from the assumption that volume is positive even in the left-handed coordinates, the conclusion is that $E$ and $D$ are polar vectors and $H$ and $B$ are axial vectors. The different starting points thus result in different conclusions as to the space reversal symmetry of $D$ and $H$. For $E$ and $B$, the same result is obtained irrespective of whether volume is positive or negative in the left-handed coordinates. The fact that $E$ is a polar vector and $B$ is an axial vector is correct even from the standpoint of vector potential $A$. Some textbooks explicitly state the symmetry of $E$ and $B$, whereas none contains a clear description of the symmetry of $D$ and $H$ except Reference 5. If $E$ and $D$ have the same symmetry and $H$ and $B$ the same symmetry, there is no reason for the existence of two kinds of fields for both electric and magnetic

Table 1 : Space reversal symmetry of various electromagnetic quantities

| Scalar  P(+) | $q$ (Charge) |
|---|---|
| Pseudoscalar  P(-) | $\rho$  (Charge density), <br> $U$ (Energy density of electromagnetic fields) <br> $\varepsilon_0$ (Permittivity of free space) <br> $\mu_0$ (Permeability of free space) |
| Polar vector  P(+) | $E$ (Electric field), $H$ (Magnetic field) |
| Axial vector  P(-) | $D$ (Electric induction or electric flux density) <br> $B$ (Magnetic induction or magnetic flux density) <br> $J$ (Current density) <br> $g$ (Momentum density of electromagnetic fields) <br> $S$ (Poynting vector) |
| Tensor  P(-) | $T_{ij}$ (Maxwell's stress) |

Source: Reference[5]

fields. Consequently, we may ask which field of $H$ and $B$ is more essential. This is the origin of $E$-$H$ and $E$-$B$ formulations. These formulations, however, are meaningless, if symmetries of $H$ and $B$ are different from each other. When volume is negative in the left-handed coordinates, the permittivity and permeability of free space are pseudoscalars. As will be discussed in the next section, pseudoscalar permittivity (or pseudoscalar permeability) can play an essential role as a quantity or an operator that is relevant to the reversal of space. If the permittivity and permeability of free space is a true scalar, on the other hand, it can bear no other meaning than that it is a mere proportional constant, which precludes the understanding of the true meaning of the universal constant.

### 3-3 Space Reversal Caused By Permittivity

Numerical values of permittivity and permeability of free space are defined as $\varepsilon_0 = 8.854187817 \times 10^{-12}$ F/m and $\mu_0 = 4\pi \times 10^{-7}$ N/A$^2$. Today, we cannot measure them because they are defined. The permittivity $\varepsilon_0$ and permeability $\mu_0$ determine the following relations between $E$ and $D$ and between $H$ and $B$ in free space (or in the air):

$$D = \varepsilon_0 \cdot E, \quad B = \mu_0 \cdot H. \tag{3}$$

When $E$ and $H$ are polar vectors with symmetry P(+) and $D$ and $B$ are axial vectors with symmetry P(-) as already discussed, the symmetry of $\varepsilon_0$ and $\mu_0$ is P(-), satisfying P(-) = P(-)·P(+). Namely, $\varepsilon_0$ and $\mu_0$ are pseudoscalars that are positive in the right-handed coordinates and negative in the left-handed coordinates, which appeared in Imai's textbook.[5] The energy density of electromagnetic fields includes both $\varepsilon_0$ and $\mu_0$ together with $E^2$ and $H^2$. Since the energy density is negative in the left-handed coordinates, $\varepsilon_0$ and $\mu_0$ are also negative in the coordinates, which can also provide an explanation for why $\varepsilon_0$ and $\mu_0$ are pseudoscalars.

The permittivity $\varepsilon_0$, the permeability $\mu_0$, and the speed of light c satisfy the relation $1/c^2 = \varepsilon_0 \cdot \mu_0$. In fact, Maxwell calculated the propagation speed of electromagnetic waves using this relation, and proposed that the visible ray of light is a kind of electromagnetic waves. The relation clearly shows that permeability is automatically determined once speed of light and permittivity are set. In this report,

I have already explained that a system of units where $c^2 = 1$ should be adopted. Since permittivity $\varepsilon_0$ and permeability $\mu_0$ are defined values today, both values can be defined as one to create simpler systems of units. Such systems of units were actually used in the past. For instance, $\varepsilon_0 = 1$ in the system of cgs-electrostatic units (cgs-esu), and $\mu_0 = 1$ in the system of cgs-electromagnetic units (cgs-emu). In the electromagnetic units, $H$ and $B$ are measured by oersted and gauss, respectively, though they show the same values in free space (or in the air). The concurrent use of electrostatic and electromagnetic units leads to $c^2 = 1$, which differs from about 300,000 km/s. For that reason the concurrent use of the two systems of units has been avoided. If $c^2 = 1$ is accepted, however, the two systems of units become concurrently usable. As mentioned above, the permittivity and permeability of free space are pseudoscalars, and they shows negative values in the left-handed coordinates. That is, $\varepsilon_0 = \mu_0 = \pm 1$, or equivalently $\varepsilon_0{}^2 = \mu_0{}^2 = 1$. Here, $\varepsilon_0$ and $\mu_0$ may be regarded as identical, having exactly the same meaning. Although either one of the two will therefore suffice for the discussion, the following deals with both to avoid confusion.

If $\varepsilon_0{}^2 = \mu_0{}^2 = 1$, equation (3) is rewritten as:

$$E = \varepsilon_0 \cdot D, \; H = \mu_0 \cdot B. \tag{4}$$

The comparison between equations (3) and (4) shows that permittivity $\varepsilon_0$ and permeability $\mu_0$ play the role of transforming a polar vector into an axial vector, and transforming an axial vector into a polar vector. We can say that $\varepsilon_0{}^2 = 1$ (and $\mu_0{}^2 = 1$) is the condition under which the original scale is maintained after transformations of "polar" →"axial" → "polar" vector, or "axial" → "polar" → "axial" vector. Since $\varepsilon_0 = +1$ refers to right-handed coordinates and $\varepsilon_0 = -1$ to left-handed coordinates, $\varepsilon_0$ may be considered an operator that distinguishes the coordinate system, or an operator that causes reversal of space.

In the universal system of units where $c^2 = \varepsilon_0{}^2 = 1$, two of four base units are eliminated, and we have only two base units. For instance, $c^2 = 1$ integrates time and length, and eliminates either meter or second. In either case, the unit of velocity disappears, and energy is measured in terms of mass. Equation $\varepsilon_0{}^2 = 1$ eliminates the ampere, and electric currents are described by a combination of the remaining two units. In such a system of units, the electromagnetic fields $E, D, H$, and $B$ have the same unit, with the only differences being in the symmetries of time reversal and space reversal. Specifically, $E$ is P(+) and T(+), $D$ is P(-) and T(+), $H$ is P(+) and T(-), and $B$ is P(-) and T(-). The four types of electromagnetic fields are all different and unique in symmetry. This is the reason why the four electromagnetic fields need to exist.

## 4 | Confusion in Textbooks on Electromagnetism

Descriptions in electromagnetism vary greatly depending on whether to start from a magnetic field generated by an electric current (or electromagnet) or by magnetic charges (or a permanent magnet). The former standpoint is referred to as *E-B* formulation, and the latter *E-H* formulation. Their characteristics are as follows:

a) *E-B* formulation:

1. Assumes neither magnetic charges nor monopoles, and the existence of monopole is denied. Magnetic field $B$ is induced by electric currents. One of Maxwell equations, div $B$ = 0, shows that magnetic field lines of $B$ are continuous.
2. Compatible with the theory of relativity. When an electron moves at the speed of $v$, $v \times B$ is the electric field that the electron feels. Further, it is $E$ and $B$ that are derived from the vector potential $A$ of an electromagnetic field.
3. It has the disadvantage that a virtual electric current must be assumed around the permanent magnet because the presence of magnetic poles "N" and "S" on both ends of the magnet is denied. Such a model is referred to as the electric current model of magnetization.

b) *E-H* formulation:

1. Assumes an magnetic dipole or a pair of magnetic charges that have opposite signs. The magnetic poles "N" and "S" on both ends of the magnet are responsible for the magnetic field $H$. It is supported by the fact that div $H \neq 0$ at the boundary of magnets; i.e., magnetic field lines of $H$ are not continuous at the boundary. One of Maxwell equations, div $B$ = 0, shows that the same amounts of magnetic charge with opposite signs should appear on the respective sides. This model is referred to as the magnetic charge model of magnetization.

2. It is the magnetic field $H$ that an electric current produces through Ampere's law rot $H = J$. Actually, the magnetic field $H$ is measured by ampere/meter.
3. It has the disadvantage of requiring the concept of a magnetic monopole, the existence of which has not been confirmed.

Many recent textbooks support the *E-B* formulation, while the ones that deal mainly with magnetism or microwaves often adopt the *E-H* formulation. The choice of the formulation may be a matter of taste, because it is not that one of formulations is correct and the other is wrong. However, different definitions and units of magnetization and susceptibility between two formulations cause the confusion far beyond a matter of taste. It is students or beginners who suffer the most from this confusion.

The confusion of *E-B* and *E-H* formulations is well recognized by authors of recent textbooks and teachers in the field of electromagnetism. There are many references to this issue.[7-9] In particular, Reference 7 contains a table that classifies some textbooks according to the formulations. Table 2 is reproduced from the classified table in Reference 7. Further, it says, "There are many textbooks on classical electromagnetism. The authors are all strongly committed to their work, several of which were written with the intention of providing a critical discussion of conventional textbooks. For example, one textbook (Hosono) insists that the *E-H* formulation is wrong and should be absolutely avoided, while another (Mizoguchi) insists that the electric current model of permanent magnet, which is the basic idea of the *E-B* formulation, is anachronistic."

The difference between the *E-B* and *E-H* formulations is caused by two different standpoints, i.e., $B$ is fundamental field or $H$ is the fundamental field. We should say again that it is a matter of choice and not a matter of which is right and which is wrong. Some textbooks that support the *E-B* formulation refer to $B$ as magnetic field, without using $H$ at all. Some textbooks on microwaves use $E$ and $H$ only. The unique textbook[5] that describe the space reversal

**Table 2 :** *E-B* and *E-H* formulations in textbooks on electromanetism

| Author (translator) | Title | Publisher | *E-B* | *E-H* | Note |
|---|---|---|---|---|---|
| Barger, Olsson (trans. by Kobayashi, Tosa) | Classical Electricity and Magnetism | Baifukan | ○ | | |
| Serway (trans. by Matsumura) | Physics for Scientists and Engineers (Electromagnetism) | Gakujutsu Tosho Shuppan | ○ | | |
| Feynman (trans. by Miyajima) | The Feynman Lectures on Physics (Electromagnetism) | Iwanami Shoten | ○ | | |
| Nagaoka, Tankei | Introductions to Physics: Q&A Exercises on Electromagnetism | Iwanami Shoten | ○ | | |
| Katsurai | Fundamental electromagnetism for science and engineering | Ohmsha | ○ | | |
| Stratton | Electro-Magnetic Theory | McGrow-Hill | ○ | | |
| Jackson | Classical Electrodynamics | Wiley | ○ | | 1 |
| Nakayama | Electromagnetism | Shokabo | ○ | ○ | 2 |
| Iida | New Electromagnetism | Maruzen | ○ | ○ | |
| Mizoguchi | -SI UNITS- Electromagnetism | Shokabo | | ○ | 3 |
| Bleaney, Bleaney | Electricity and Magnetism | Oxford | | ○ | |
| The Institute of Electrical Engineers of Japan | Exercises on Electromagnetism | Ohmsha | ○ | ○ | 4 |
| Halliday, Resnick, Walker (trans. by Nozaki) | Fundamentals of Physics, III: Electromagnetism | Baifukan | ○ | | |
| Goto | Comprehensible Electromagnetism | Kodansha | | ○ | |
| Kumagai, Arakawa | Electromagnetism | Asakura | | ○ | |
| Takahashi | Physic Selection: Electromagnetism | Shokabo | | ○ | |
| Shimoda, Chikazumi | Exercises in College: Electromagnetism | Shokabo | | ○ | |
| Tokai Univ | Physics: Electromagnetism | Tokai Univ. Pres | ○ | | |
| Hirose | Physics One Point: *E* and *H*, *D* and *B* | Kyoritsu Shuppan | | ○ | |
| Hosono | Meta-electromagnetism | Morikita Publ. | ○ | | 5 |
| Kozuka | Electricity and magnetism: Its physical images and details | Morikita Publ. | | ○ | |
| Landau, Lifshitz (trans. by Inoue, Yasukouchi, Sasaki) | Electromagnetism | TokyoTosho | ○ | | 1 |
| Suematsu | Electromagnetism | Kyoritsu Shuppan | ○ | | |
| Sunakawa | Electromagnetism | Iwanami Shoten | ○ | | 6 |
| Slater, Frank | Electromagnetism | McGrowHil | ○ | | |
| Nagaoka | Introductions to Physics: Electromagnetism I, II | Iwanami Shoten | ○ | | |
| Murakami | Electromagnetism | Maruzen | | ○ | |
| Sommerfeld (trans. by Ito) | Electromagnetism | Kodansha | ○ | | |
| Purcell | Berkeley physics coures vol. 2 Electricity and magnetism | McGrowHill | ○ | | 1 |
| Ota | Fundamentals of Electromagnetism I, II | Springer Japan | ○ | | 7 |

(Notes 1 to 6 are quoted from Reference 7)

1) This old textbook is written in Gaussian system of units (cgs-emu).

2) Supports both *E-B* and *E-H*. Leaning toward *E-H*.

3) Closer to the standard theoretical development on *E-H* model, but rejects the idea of a "magnetic charge" even as a virtual entity. Takes the strict stance that magnetic substance is essentially a set of magnetic moments.

4) A textbook that deals equally with both *E-B* and *E-H* formulations.

5) A unique book that is critical of *E-H* formulation. Claims that *E-H* formulation is wrong since a single moving magnetic charge violates relativistic invariance.

6) A textbook that is based on a compromise between *E-B* and *E-H*. Uses a pole model for magnetic substance.

7) Claims that *E-H* formulation lacks theroretical basis since *E* and *B* are inseparable under the theory of relativity.

Source: Reference[7]

symmetry of **H** and **D** takes a stance closer to the D-B formulation, because it states that "the most fundamental quantities in describing electromagnetic fields are **D** and **B**."

Which quantities are more fundamental is only a subjective matter. The important thing is to discuss the space reversal symmetry of electromagnetic fields because it is confused at present, although the time reversal symmetry of electromagnetic fields may not be confused. I feel that almost all textbooks lack sufficient discussion of the space reversal symmetry of electromagnetic fields. If **E** and **H** are polar vectors and **D** and **B** axial vectors, then the necessity of the four types of fields is concluded. If, on the other hand, **E** and **D** are polar vectors and **H** and **B** are axial vectors, there is no reason for the existence of two types of fields for both electric and magnetic fields, and the conclusion must be that there are fundamental fields and secondary fields. It causes unnecessary controversy as to whether E-B or E-H formulation.

## 5  On Charge Reversal Symmetry

All electrons and all protons have the same amount of electric charge of $1.6021773 \times 10^{-19}$ Coulombs, which is called a unit charge or an elementary charge. However, it is not known exactly whether such a fundamental unit of magnetic charge exists, or whether a magnetic charge can exist by itself. If magnetic charges exist, it is certain that magnetic field lines start from positive charge and end with negative charge in the same way as electric field lines. So, the question is thus whether a magnetic field line is continuous or has a start point and an end point. Figures 4(a) and 4(b) shows magnetic field lines of **B** and **H**, respectively, in and around the spherical permanent magnet. While textbooks often show magnetic field lines of a bar magnet, there is no difference between a spherical and a bar magnets except that the magnetic field lines inside a bar magnet are somewhat more complicated.

Given $\mu_0 = 1$, the magnetic field lines of **B** and those of **H** coincide with each other outside the magnet, although the lines of **B** are denser than those of **H** inside the magnet. An essential difference between the magnetic field lines of **B** and **H** is that the lines of **H** are not continuous at the surface of the magnet and have a start point and an end point (Figure 4(b)) there, while lines of **B** are continuous and have no start point or end point (Figure 4(a)). This suggests that magnetic charges of N and S poles exist as the source of the magnetic field **H**. Here, we note that **B** and **H** show different directions inside the magnet. Meanwhile, some textbooks of E-B formulation do not accept the existence of magnetic charge, often denying the reality of magnetic field lines of **H**.

Dirac, who is famous for predicting antiparticles, argued based on the quantum mechanics that if a magnetic monopole exists, its magnetic charge $g$ and electron charge $e$ have to satisfy the following equation:

$g$ (monopole charge) =
$h$ (Planck constant) $\cdot c$ (speed of light)/$e$
(electron charge).               (5)

The Planck constant $h$ is a universal constant characteristic to quantum mechanics, and $c$ is the speed of light as previously mentioned. According to Dirac, a line with phase singularity, which is call Dirac string, extends from a magnetic monopole to infinity (Figure 5(a)) like a filament appearing in the center of a vortex. The discovery of a particle with monopole charge, i.e., a particle with either only a N or S pole, would answer the question of "why every electron in the universe should have exactly the same electric charge"[10] on the basis of equation (5). Many scientists have been searching but have yet to find a particle with monopole charge.

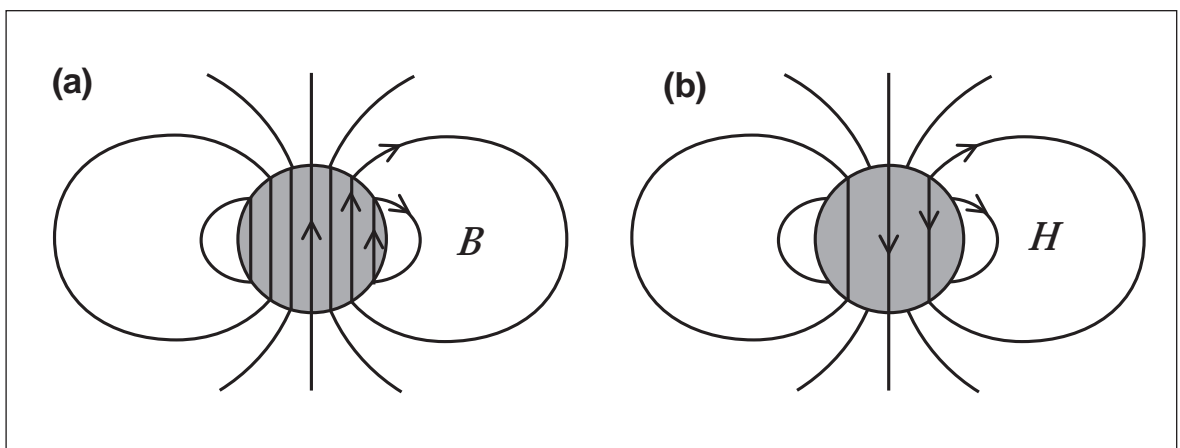

**Figure 4 :** The field lines of (a) magnetic induction **B** and (b) magnetic field **H** inside and outside a spherical magnet

If the electron charge $e$ is replaced with the charge $2e$ of a Cooper pair, equation (5) coincides with the equation of the flux quantum in a superconductor. A magnetic field penetrates a superconductor not uniformly but in a quantized form, which is referred to as the flux quantum (Figures 5(b) and 5(c)). Flux quanta have been actually observed and ascertained to exist. The immediate cause of the quantization of the magnetic field is a persistent current flowing around the magnetic field lines. At the center of the vortical current, there is a string where superconductive phase cannot be defined. The flux quantum is thus sometimes called a vortex string, which is nothing but a Dirac string excepting the difference of the charge $e$ or $2e$. The magnetic monopole given by Dirac and the flux quantum share exactly the same theoretical basis, with the only difference being whether there are one or two electrons. In view of this, we can assume that a flux quantum is accompanied by magnetic monopoles at the respective ends. Even in such a case, the field lines of the magnetic induction $B$ are continuous (Figure 5(b)), and the magnetic charges serve not as the source of the magnetic induction $B$ but as the source of the magnetic field $H$ (Figure 5(c)). In that sense, Figures 5(b) and 5(c) correspond to Figures 4(a) and 4(b), respectively.

In Maxwell's equations, doubling the electric quantity results in double the magnetic fields $H$ and $B$. If a magnetic monopole (or it's pair) exists as the source of $H$, the magnetic monopole charge $g$ is expected to be proportionate to the charge $e$ of the electron:

$g$ (monopole charge) =
$c$ (speed of light) $\cdot$ $e$ (electron charge).    (6)

In this equation, the speed of light $c$ ($= \pm 1$) is multiplied for the reason that the electric field $E$ and the magnetic field $H$ differ in time reversal symmetry. Equation (5), where monopole charge is inversely proportional to electron charge, and equation (6), where they vary in direct proportion, seem to be contradictory but actually need to hold at the same time. Equations (5) and (6) immediately yields $e^2 = g^2 = 1$, showing that both electron and monopole charges have unique value except for it's positive or negative sign. In this discussion, we assumed $h = 1$ and $c^2 = 1$. In the ordinary MKSA system of units, we will obtain the magnetic monopole charge of $\pm 4.14 \times 10^{-15}$ Webers, which is twice the value of magnetic flux quantum in a superconductor.

The charge of an electron was historically defined to be negative. In $e^2 = 1$, $e = -1$ represents an electron and $e = 1$ represents a positron or an antiparticle of the electron. In semiconductors, $e = -1$ represents an n-type carrier or electron, and $e = 1$ a p-type carrier or hole. As can be seen, elementary charge $e$ ($= \pm 1$) is a quantity (or operator) that distinguishes a particle from an antiparticle, and can thus be regarded as an operator that is relevant to C-symmetry (charge symmetry or particle reversal symmetry) within CPT symmetry. There are quantities whose sings (or directions) are reversed when a particle is reversed to an antiparticle or when all particles are reversed to antiparticles. Examples of these quantities are electromagnetic fields, $\boldsymbol{E}$, $\boldsymbol{D}$, $\boldsymbol{H}$, and $\boldsymbol{B}$. We will denote this particle reversal symmetry by C(-). In contrast, the symmetry where the sign does not change will be denoted by C(+). Time derivatives, spatial derivatives, permittivity, and permeability show C(+) symmetry. By help of notations C(+) and C(-), the symmetry of equations can be easily discussed in the same way as in the case of T and P. We will find that Maxwell's equations conserve the particle reversal symmetry.

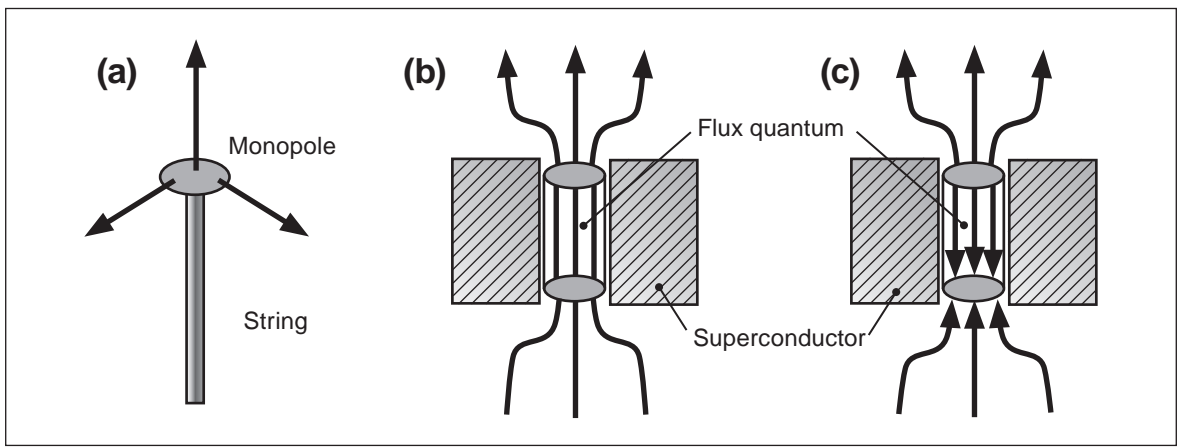

**Figure 5 :** (a) Dirac's magnetic monopole, (b) the field lines of a flux quantum in terms of **B**, and (c) those in terms of **H**. The flux quantum can be considered to have monopoles on both ends

## 6    Particle-Wave Duality

I have explained that the speed of light $c$ is relevant to time reversal, permittivity $\varepsilon_0$ to space reversal, and elementary charge $e$ to particle reversal. Thus, universal constants are (eigenvalues of) operators closely associated with CPT symmetry, provided that $c^2 = \varepsilon_0{}^2 = e^2 = 1$. The remaining universal constant is the Planck constant $h$ ($= 6.62606896 \times 10^{-34}$ joule-seconds), which appears in quantum mechanics.

The most important conclusion of quantum mechanics is that an electron has two mutually contradictory properties of a particle and of a wave. It is the Planck constant $h$ that links the particle and wave. This constant was found by Planck through spectrum analyses of black body radiation, pioneering quantum mechanics. Later on, Einstein and de Broglie derived the significant relations:

$$E = h \cdot \nu \text{ and } \boldsymbol{p} = h \cdot \boldsymbol{k}, \qquad (7)$$

which provided photonic quanta by Einstein and matter waves by de Broglie. De Broglie himself did not consider electrons to be waves, but assumed the electron to be on top of the wave and formulated the foregoing relations. In equations (7), $\nu$ is frequency and $\boldsymbol{k}$ is wave vector (= reciprocal of the wavelength), both of which are quantities of waves. $E$ and $\boldsymbol{p}$ are energy and momentum of a particle or quantum, both of which are quantities that describe the state of a particle. Consequently, equations (7) show that wave-describing quantities multiplied (operated) by the Planck constant $h$ make particle-related quantities.

If we assume $h^2 = 1$, or employ such a system of units, then equations (7) give:

$$\nu = h \cdot E \text{ and } \boldsymbol{k} = h \cdot \boldsymbol{p}. \qquad (8)$$

Equations (8) show that particle-related quantities multiplied (operated) by the Planck constant $h$ make wave-describing quantities. The Planck constant $h$ $= \pm 1$ can thus be regarded as a quantity or operator that transforms a wave and a particle into each other. $h^2 = 1$ is also the condition that prevents a change in scale (such as energy and momentum) under transformations. Since $h = \pm 1$, energy and frequency have the same unit, and momentum and wave vector have the same unit.

In the presence of equations (5) and (6), however, the quantity $h$ is automatically determined once $c^2 = \varepsilon_0{}^2 = e^2 = 1$ is given. Thus, even if we let $c^2 = \varepsilon_0{}^2 = e^2 = h^2 = 1$, three units disappear out of the four units in the MKSA system of units, leaving just one. The remaining unit may be any one of the four. For example, if the unit of time is left, all quantities are measured by the unit of time or by no units.

The assumption that $h^2 = 1$ implies that there is a negative Planck constant ($h = -1$) as well as the positive Planck constant ($h = 1$). The positive and negative values can be associated with the symmetry of particles and antiparticles. A reasonable explanation can be provided by assigning the Planck constant $h$, the imaginary unit $i$, and the wave-related quantities $\nu$ and $\boldsymbol{k}$ as C(-) and assigning the particle-related quantities $E$ and $\boldsymbol{p}$ as C(+). For example, equations (7) and (8) have particle symmetries C(+) = C(-) $\cdot$ C(-) and C(-) = C(-) $\cdot$ C(+), respectively. Since the symmetry of energy $E$ is C(+), energy $E$ is always positive for both particles and antiparticles, while frequency of the positron (or antiparticle) waves is negative. From equations (7) and (8), the application of the Planck constant to a wave produces a particle, and the application of the Planck constant to a particle produces a wave. What is actually occurring has yet to be clarified, however. I would venture to say that a particle is not other than a wave, and a wave is not other than a particle, because a particle is immediately transformed into the wave and a wave is immediately transformed into the particle ("emptiness is not other than form, form too is not other than emptiness; form is emptiness, emptiness is form"— *The Heart Sutra*). Each individual electron, repeating such a process of reincarnation, would make existence eternal, but why a single electron survives eternally is not yet clearly understood.

# 7 | Conclusion

Having started from the questions of why the universal constants in physics, such as the speed of light $c$, permittivity $\varepsilon_0$ or permeability $\mu_0$ of free space, the electron's charge $e$, and the Planck constant $h$, are constant and why such values should be taken, I have deduced that universal constants are closely associated with CPT symmetry, i.e., reversal symmetries of time, space, and particle. I have also pointed out that the system of units where $c^2 = \varepsilon_0^2 = e^2 = h^2 = 1$ is essential to the discussion.

Some people already had these kinds of questions. Those who claim that the theory of relativity is wrong, for example. Although it is absolutely clear that their argument is wrong and the theory of relativity is correct, it should be noted that what they are concerned with is the principle of the constancy of the speed of light, or the question of why the speed of light is constant. Specialists have so far not answered this question squarely. It may be because they have considered its constancy to be natural and beyond question, or perhaps they have regarded it as an axiom that is unexplainable and unnecessary to be explained. Some scientists, however, raised the question of why every electron in the entire universe has exactly the same electric charge. Dirac and Yukawa were among them. In order to solve this question, Dirac introduced the existence of a unit magnetic charge or monopole through a duality of electric and magnetic fields in Maxwell's equations. This was only half-successful, however, in providing a complete solution to the question. Electrons have a magnetic moment called spin, which is well known to be the source of a permanent magnet. Recently, the properties of spin have been utilized in electronics, and the term spintronics has become prevalent. Electron spin or it's magnetic moment is often described as a rotation model of an electron. The electron itself, however, is known to be a point charge with no dimensions, and the rotation of a point charge will not produce any magnetic moment. A valid model of spin has not been constructed yet. Likewise, what we consider obvious is often not yet fully understood. Meanwhile, there are questions that many people have been aware of, such as the question of how to interpret the particle-wave duality. This question was relevant to the Bohr-Einstein debates about the probabilistic interpretation of wave functions, and is still discussed as a measurement problem.

Such "naive but reasonable" questions are essential to a new revolution. Undoubtedly, there are also important questions that we are still unaware of. The question of whether the volume in the left-handed coordinates is positive or negative was raised by Maxwell some 130 years ago, but has been neglected thereafter. This question, that has not been fully resolved, is responsible for the confusion of *E-B* or *E-H* formulation in the modern textbooks on electromagnetism. Since *E-H* formulation may need magnetic charges, the confusion may be relevant to the question of how Dirac's monopole (or it's pair) should be dealt with in Maxwell's equations.

Finally, I want to say that it is important to discover and unearth naive but reasonable questions. Such questions are, however, less likely to be found among specialists in their fields. Furthermore, we have to recall the history such as "what many people consider as a fiction can turn out to be true, like Planck's quantum and de Broglie's matter wave."[11] One possible approach may be to offer a prize for finding such reasonable questions, but not for solving the questions. This will be a new version of prize essay that made a considerable contribution to mathematics and basic science in 18th- and 19th-century Europe. In any case, scientists who engage in tenacious efforts aimed at finding novel and reasonable questions and at pursuing such basic problems are increasingly significant. It is of extreme importance to develop such research environments and human resources in Japan.

**Acknowledgement**

This report has benefited from information and comments on the confusion in textbooks on electromagnetism offered by Prof. Endo, along with permission to quote from his website. Professor Yasue also provided information and comments on the interpretation of electromagnetic fields and the monopole from the standpoint of elementary particle theory. I would like to express my deepest gratitude to them.

## References

[1] Susumu Yamaga, "Science of Space," : http://www.s-yamaga.jp/nanimono/uchu/jikokutokoyomi-02.htm

[2] National Institute of Information and Communications Technology, press release : http://www2.nict.go.jp/pub/whatsnew/press/h20/080912/080912-1.html

[3] Chuji Adachi, Vector Analysis, pp. 32-3, Baifukan.

[4] James Clerk Maxwell, A Treatise on Electricity and Magnetism; 1st ed. (1873), 2nd ed. (1881), 3rd ed. (1891) by Clarendon Press. The 3rd edition is republished by Dover Publications, Inc. Also see : http://rack1.ul.cs.cmu.edu/is/maxwell1/

[5] Isao Imai, Mathematics in Classical Physics, Iwanami Series of Lectures on Applied Mathematics, Iwanami Shoten.

[6] Shigeo Kobata, Thus were Created the Electromagnetic Units, Kougakusha.

[7] On Textbooks of Electromagnetism, http://teamcoil.sp.u-tokai.ac.jp/classes/EM1/Unit/index.html (the website by Prof. Endo, Tokai University), to which descriptions of Reference 5 have been added after our discussion.

[8] EMAN's Physics: Electromagnetism : http://homepage2.nifty.com/eman/electromag/em_unit.html

[9] E-B formulation and E-H formulation, Japanese Wikipedia.

[10] Hideki Yukawa, Yasuhisa Katayama, Hideo Fukutome, Elementary Particles, Iwanami Shinsho, 1961 (2nd ed. 1969).

[11] Takehiko Takabayashi, History of Development of Quantum Theory, Chikuma Shobo, 2002.

## Profile

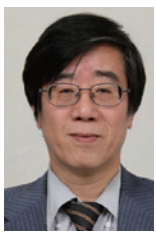

**Tsuneo ICHIGUCHI**
Leader of Information and Communications Research Unit, STFC
http://www.nistep.go.jp/index-j.html

Ph.D. in Physics. Specialized in physics of semiconductor, superconductor, and magnetism. Engaged in research, primarily on measurement with electromagnetic waves, in American university and Japanese electric appliance manufacture.

(Original Japanese version: published in June 2009)

# SUPPLEMENT : Symmetries of various equations

We believe that this paper shows new concepts in fundamental physics, although it was written for wider audiences who are nonspecialists. Some sections of the present paper would be unnecessary for specialists, and detailed descriptions will be insufficient. Because writing equations is not usually allowed in Science & Technology Trends, we supply some details with related equations in this supplement. Using symbols $P(+/-)$, $T(+/-)$, and $C(+/-)$ for reversal symmetries, we find symmetries of Maxwell equations:

$$\boldsymbol{\nabla}\times \boldsymbol{E} + \mu_0 \ \partial t \ \boldsymbol{H} = -\boldsymbol{j}_M = 0 \,, \quad \text{(A1)}$$
$$P(-)P(+) + P(-)P(+)P(+) = P(-)$$
$$T(+)T(+) + T(+)T(-)T(-) = T(+)$$
$$C(+)C(-) + C(+)C(+)C(-) = C(-)$$

$$\boldsymbol{\nabla}\times \boldsymbol{H} - \epsilon_0 \ \partial t \ \boldsymbol{E} = \boldsymbol{j} \,, \quad \text{(A2)}$$
$$P(-)P(+) - P(-)P(+)P(+) = P(-)$$
$$T(+)T(+) - T(+)T(-)T(+) = T(-)$$
$$C(+)C(-) - C(+)C(+)C(-) = C(-)$$

$$\epsilon_0 \ \boldsymbol{\nabla}\cdot \ \boldsymbol{E} = \rho \,, \quad \text{(A3)}$$
$$P(-)P(+)P(+) = P(-)$$
$$T(+)T(+)T(+) = T(+)$$
$$C(+)C(+)C(-) = C(-)$$

$$\mu_0 \ \boldsymbol{\nabla}\cdot \ \boldsymbol{H} = \rho_M = 0 \,, \quad \text{(A4)}$$
$$P(-)P(+)P(+) = P(-)$$
$$T(+)T(+)T(-) = T(-)$$
$$C(+)C(+)C(-) = C(-)$$

where we put $\boldsymbol{j}_M$ and $\rho_M$ because the value 0 has no symmetries. we should note that $\boldsymbol{H}$ is a polar vector (or polar vector field) and $\boldsymbol{D}(=\epsilon_0\boldsymbol{E})$ is an axial vector. If we define

$$\rho_M(x) = g \sum_i \lim_{a\to 0}\{\delta(x_i+a-x) - \delta(x_i-x)\} \,, \quad \text{(A5)}$$

$$\boldsymbol{j}_M(x) = g \sum_i \boldsymbol{v}_i \lim_{a\to 0}\{\delta(x_i+a-x) - \delta(x_i-x)\} \,, \quad \text{(A6)}$$

then we can easily obtain the continuity equation:

$$\frac{\partial \rho_M}{\partial t} + \boldsymbol{\nabla}\cdot\boldsymbol{j}_M = 0 \,. \quad \text{(A7)}$$

Even though $\rho_M \sim 0$ and $\boldsymbol{j}_M \sim 0$, Eq.(A7) may still have a meaning when we consider the singularity like a $\delta$-function. Here, $\Sigma_i$ is the summation of particles.

Since energy density $U$ is P(-) when the volume is negative in the left-handed coordinates, it is clear that both $\epsilon_0$ and $\mu_0$ are P(-):

$$2\,U = \epsilon_0 \ \boldsymbol{E} \cdot \boldsymbol{E} + \mu_0 \ \boldsymbol{H} \cdot \boldsymbol{H} \,. \quad \text{(A8)}$$
$$P(-) = P(-)P(+)P(+)+P(-)P(+)P(+)$$
$$T(+) = T(+)T(+)T(+)+T(+)T(-)T(-)$$
$$C(+) = C(+)C(-)C(-)+C(+)C(-)C(-)$$

The violation of time-reversal symmetry in Ohmic low is expressed as follows:

$$\boldsymbol{j} = \rho \ \boldsymbol{v} = \sigma \ \boldsymbol{E} \,, \quad \text{(A9)}$$
$$P(-) = P(-)P(+) = P(-)P(+)$$
$$T(-) = T(+)T(-) \neq T(+)T(+) \,; \text{ broken}$$
$$C(-) = C(-)C(+) = C(+)C(-)$$

The symmetries of de Broglie-Einstein relation are:

$$\varepsilon = h \ \nu \ (+ \ e \ \varphi_0) \,, \quad \text{(A10)}$$
$$C(+) = C(-)C(-)+C(-)C(-)$$
$$T(+) = T(+)T(+)+T(+)T(+)$$

$$\boldsymbol{p} = h \ \boldsymbol{k} \ (+ \ e \ \boldsymbol{A}) \,. $$
$$C(+) = C(-)C(-)+C(-)C(-)$$
$$T(-) = T(+)T(-)+T(+)T(-)$$
$$\text{all } P(+)$$

Thus, energy is always positive. The fine-structure constant $\alpha$ is

$$2\alpha = (e^2)(\epsilon_0^{-1})(h^{-1})(c^{-1}) \,. \quad \text{(A11)}$$
$$P(-) = P(+) \ P(-) \ P(+) \ P(+)$$
$$T(-) = T(+) \ T(+) \ T(+) \ T(-)$$
$$C(-) = C(+) \ C(+) \ C(-) \ C(+)$$

---

[1] This paper should be cited also as "SCIENCE & TECHNOLOGY TRENDS, QUARTERLY REVIEW No.33, p.25-40" (http://www.nistep.go.jp/achiev/ftx/eng/ stfc/stt033e/qr33pdf/STTqr3302.pdf)

[2] Address: National Institute of Science and Technology Policy (NISTEP), Ministry of Education, Culture, Sports, Science and Technology (MEXT), Kasumigaseki, Tokyo, Japan 100-0013 e-mail: ichiguchi@nistep.go.jp